\documentclass[twocolumn,preprintnumbers,amsmath,amssymb]{revtex4}
\usepackage{graphicx}% Include figure files
\usepackage{dcolumn}% Align table columns on decimal point
\usepackage{bm}% bold math
\usepackage{epsfig}
\usepackage{amssymb}
\usepackage{psfrag}
\usepackage{natbib}
\usepackage{epstopdf}

\begin{document}

%\preprint{APS/123-QED}

\title{Capstan friction model for DNA ejection from bacteriophages}% Force line breaks with \\

\author{Sandip Ghosal}
\altaffiliation[permanent address: ]{Department of Mechanical Engineering \& 
Engineering Sciences and Applied Mathematics, 
Northwestern University, Evanston, IL 60208, USA}%Lines break automatically or can be forced with \\
\email{s-ghosal@northwestern.edu}
\affiliation{%
University of Cambridge, Department of Physics, Cavendish Laboratory,
JJ Thomson Avenue, Cambridge CB3 0HE, UK 
}%

%\affiliation{

\date{\today}% It is always \today, today,

\begin{abstract}
Bacteriophages infect cells by attaching to the outer membrane and injecting their DNA into the cell. 
The phage DNA is then transcribed by the cell's transcription machinery.
A number of physical mechanisms by which DNA can be translocated from the phage capsid into the cell 
have been identified. A fast ejection driven by the elastic and electrostatic potential energy
of the compacted DNA within the viral capsid appears to be used by most phages, at least to initiate infection. 
In recent {\em in vitro} experiments, the speed of DNA translocation from a $\lambda$ phage 
capsid has been measured as a function of ejected length over the entire duration of the event.
Here a mechanical model is proposed that is able to explain the observed 
dependence of exit velocity on ejected length, and that is also consistent with the accepted picture of the 
geometric arrangement of DNA within the viral capsid.
\end{abstract}

\pacs{87.15.hj, 87.16.ad, 82.37.Rs,87.80.Fe} % PACS, the Physics and Astronomy
\keywords{bacteriophage, DNA ejection, capstan problem}%Use showkeys class option if keyword
\maketitle

%\section{\label{sec:level1}First-level heading:\protect\\ The line
%break was forced \lowercase{via} \textbackslash\textbackslash}
%\section{Introduction}
\begin{figure}[t] %  figure placement: here, top, bottom, or page
   \centering
   \includegraphics[width=0.5\textwidth]{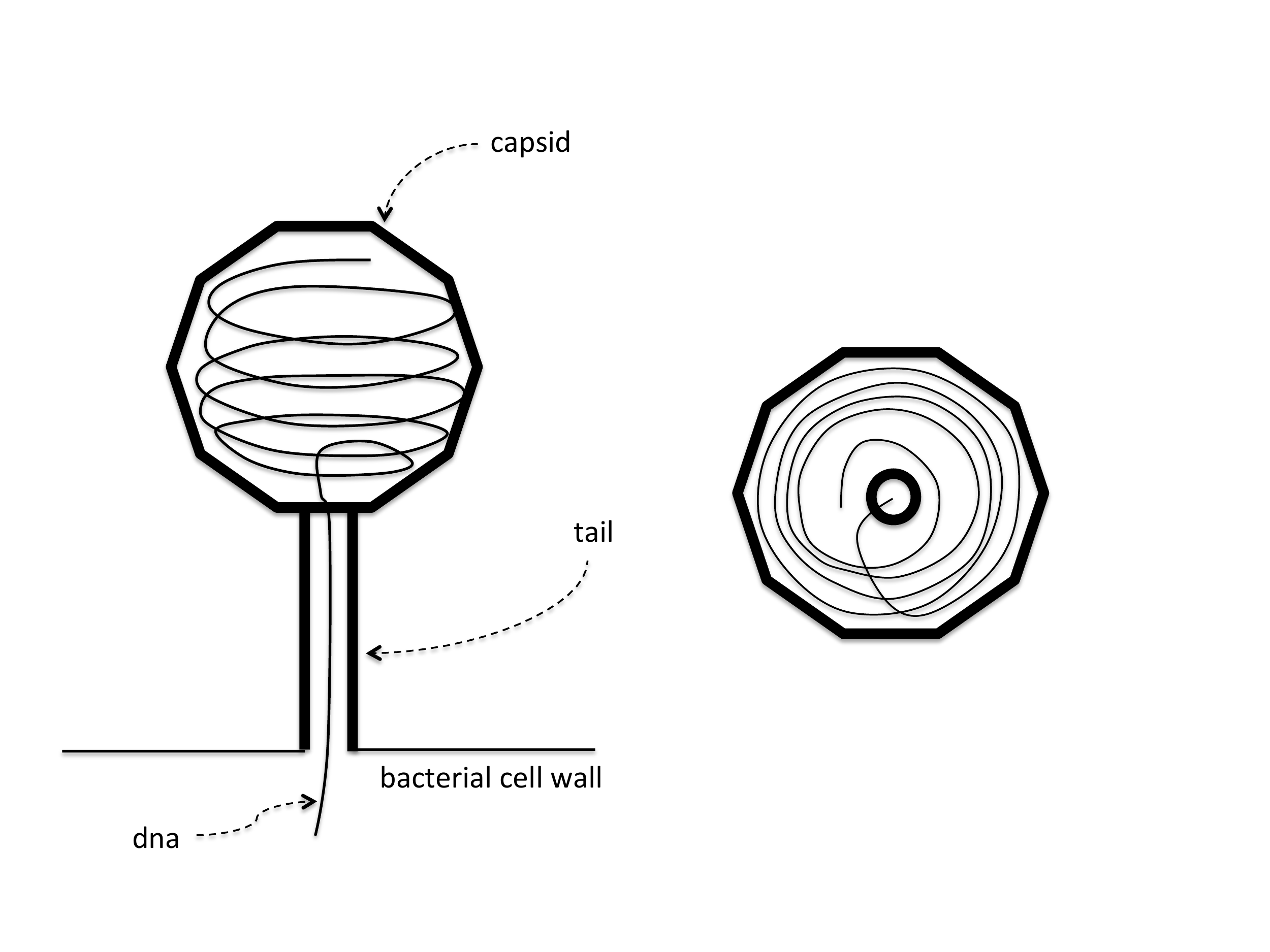} 
   \caption{Sketch showing the arrangement of the viral DNA in the capsid. The right panel shows 
   a sectional view when the point of observation is on the central axis.
    The DNA enters the capsid through the tail opening (central circle on the right panel) and is forced to curve on encountering the 
   wall. It then wraps around the central axis in helical coils of decreasing radius.}
   \label{fig:sketch1}
\end{figure}
The physics of semi-flexible polymers under confinement is a subject of great interest 
with wide applications in biology and soft matter. Viruses are an interesting example of such an application.
Bacteriophages are viruses that infect bacteria. During infection, the virus capsid itself remains outside 
the bacterium, only the DNA is injected into the cell~\cite{hershey_independent_1952}. Several physical mechanisms 
by which bacteriophages inject DNA into their hosts have been identified. The specific mechanism varies by phage 
type and stage of the infection process~\cite{molineux_fifty-three_2006,inamdar_dynamics_2006,grayson_is_2007}. 
A fast ejection on the timescale of seconds can be 
achieved simply by the release of elastic and electrostatic energy of the coiled up DNA confined within 
the viral capsid. It has been shown in {\em in vitro} studies that $\lambda$ phage can eject its entire 48.5 kbp genome by this 
mechanism in a few seconds~\footnote{Interestingly, very recent work appears to indicate that even the $\lambda$ phage infection 
mechanism may work differently {\em in vivo}~\cite{vanvalen_single-molecule_2012}}.
In other phage species, this ``syringe'' mechanism may initiate the infection, which is then completed 
through other pathways~\cite{molineux_no_2001}. 

A schematic diagram of the DNA ejection process is shown in Fig.~\ref{fig:sketch1}. 
If $F(N)$ denotes the free energy of the compacted DNA, then $-F^{\prime}(N)/b$ is the force driving the ejection, 
where $N$ is the number of base pairs  of DNA 
confined within the capsid and $b \approx 0.31$ nm is the distance between base pairs. The free energy 
of the part of the DNA that is already outside the viral capsid is neglected as it is 
much smaller than the contribution from the confined part. Since inertia is completely negligible at such small scales, 
this driving force must be balanced by a frictional force, $v m(N)$, where $v$ is the DNA translocation 
velocity and $m$ is a frictional coefficient (mobility) which may in general depend on $N$. 
The linear dependence 
on the velocity is a consequence of the fact that at low Reynolds numbers, the fluid equations reduce 
to the linear Stokes equation. Thus, if the functions $F(N)$ and $m(N)$ were known, the translocation 
velocity could be determined from 
\begin{equation} 
- F^{\prime}(N)/b = v m(N).
\label{force_balance}
\end{equation}

The function $F(N)$ can be calculated analytically~\cite{purohit_mechanics_2003,purohit_forces_2005} on the presumption that the DNA is 
packed in the 
capsid as helical coils wound in successive layers of decreasing radii in an ``inverse spool'' arrangement (Fig.~\ref{fig:sketch1}).
It has been shown that such an ordered packing does indeed minimize the free energy 
of the system. In Brownian dynamics simulations~\cite{kindt_dna_2001}, as the chain is gradually introduced into the capsid, 
it is observed to arrange itself in a donut shape. As $N$ becomes larger, it winds into a spool from the outside in. When the 
internal diameter of the spool shrinks sufficiently, the DNA stops wrapping in helical coils and simply forms loose 
turns parallel to the spool axis.  The most direct evidence for the helical conformation of DNA within the capsid has come from 
 imaging of DNA inside T7~\cite{cerritelli_encapsidated_1997} 
and T4~\cite{olson_structure_2001} phages by 
cryo-electron microscopy and of  P22 and $\lambda$ phages by X-ray 
diffraction~\cite{earnshaw_dna_1977}. The images 
reveal multiple layers of helical coils of DNA wound around the 
central axis.

The analytical formula for the free energy $F(N)$ has been subjected to direct 
experimental tests by a rather ingenious method. Real cells exert significant osmotic 
pressures that are believed to resist DNA insertion by the phage. It has been possible to 
actually ``stall'' the DNA translocation in {\em in vitro} experiments by raising the osmotic 
pressure in the solution outside the capsid. In such cases, the translocation stops with 
an amount $N=N_{0}$ of DNA still remaining within the capsid, where $N_{0}$ is 
 given by 
\begin{equation} 
- F^{\prime}(N_{0})/b = \Delta p_{os} \pi a^{2}.
\label{osmotic_balance} 
\end{equation} 
Here $\Delta p_{os}$ is the difference in osmotic pressures between the outside and inside the 
capsid and $a$ is the DNA radius.
The quantities $\Delta p_{os}$ and $N_{0}$ can be directly measured for a range of values: from 
$\Delta p_{os} = 0$ to a value that is high enough that none of the DNA is ejected. Thus,
$F(N)$ can be determined experimentally and compared to analytical models based on the elastic and 
electrostatic energy of a helically coiled semi-flexible rod. 
This has been done~\cite{evilevitch_measuring_2004,grayson_effect_2006} and the theoretical 
models for  $F(N)$ were found to agree very well with experimental data. Further evidence for the 
model comes from studying an event that occurs much later in the infection process: the packaging 
of viral DNA into the capsid prior to lysis of the host cell and release of the phages into the environment.
In an {\em in vitro} experiment, Smith {\it et al}~\cite{smith_bacteriophage_2001} attached an optical 
 bead to a DNA strand being actively packaged into the $\phi$29 bacteriophage capsid by 
 molecular motors. By applying a variable force to the optical bead and observing the 
 changes in the packing rate, they concluded that the elastic force against which the 
 molecular motors must work, increases rapidly with the amount of DNA inside the 
 capsid. This is again consistent with the rapidly decreasing nature of the function $F(N)$.
 
\begin{figure}[htbp] %  figure placement: here, top, bottom, or page
  \centering
\includegraphics[width=0.4\textwidth]{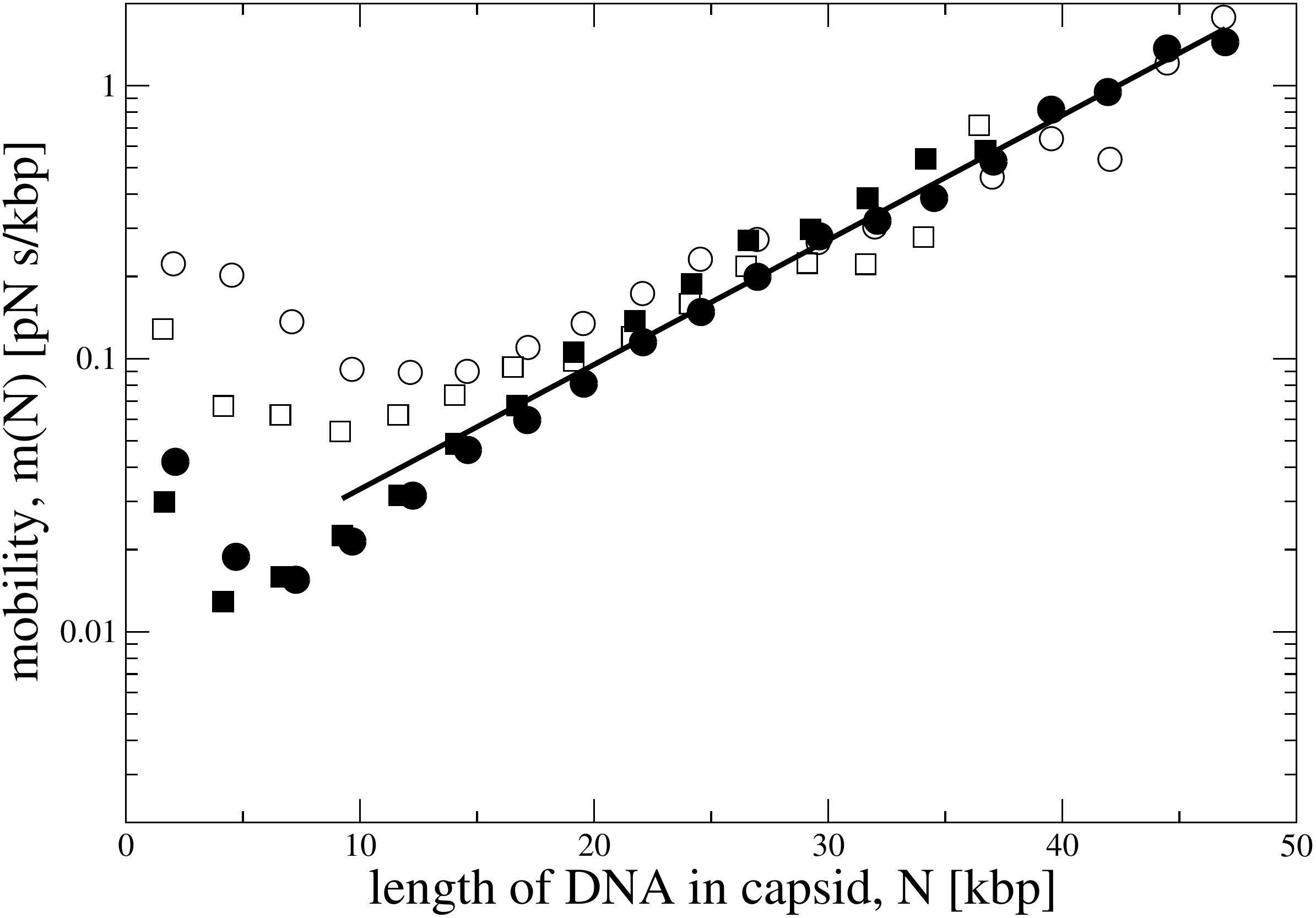}
   \caption{Dependence of mobility on DNA length remaining in capsid. 
   Measured values of  m(N) from Fig. 5 in the paper by 
   Grayson {\it et al.}~\cite{grayson_real-time_2007}
   shown with the fitted exponential, equation~(\ref{visc_drag_exp}). The data are for
   $\lambda$cI60 in NaCl buffer (solid circle), $\lambda$b221 in NaCl buffer (solid square),
   $\lambda$cI60 in MgSO$_{4}$ buffer (open circle) and   $\lambda$b221 in MgSO$_{4}$ buffer
   (open square). Error bars are not shown.}
     \label{fig:plot}
\end{figure}
Unfortunately, no such theoretical result for the function $m(N)$ in equation~(\ref{force_balance}) is 
available. However, if $v$ is measured experimentally, then using equation~(\ref{force_balance}) 
the function $m(N)$ can be obtained. This was done in a recent experiment by 
Grayson {\it et al.}~\cite{grayson_real-time_2007} where 
$\lambda$ phages were immobilized on a surface and induced to eject their DNA by exposing 
them to cell surface proteins to which the phage would normally attach and 
trigger the ejection event in the native state. The ejected DNA 
was visualized by a fluorescent dye and the ejection process recorded by 
video microscopy.  The data from the experiment  (Fig.~5 in their paper) 
 is reproduced in Fig.~\ref{fig:plot} from which certain conclusions may be drawn about 
the origin of the frictional resistance encountered by the DNA during the ejection process. 

Frictional resistance to the motion of DNA can arise either from friction within the 
phage tail or from movement of the DNA outside or within the capsid. In the former case, 
 $m(N)$ would be independent of $N$.
Fig.~\ref{fig:plot} shows that $m(N)$ becomes approximately independent of $N$ 
only towards the very end of the translocation process after 80 to 90 percent of the 
DNA has already been ejected. This is consistent with the following 
simple estimate of the frictional drag that is obtained if the DNA is taken as a cylinder 
of radius $a \sim 1$ nm, concentric with a slightly larger cylinder (the phage tail) of 
radius $a_{*} \sim 1.5$ nm:
\begin{equation} 
f_{\text{drag}} = \mu \frac{v}{a_{*}-a} 2 \pi a L
\label{visc_drag_cyl}
\end{equation} 
where $L \sim 150$ nm is the length of the tail and $\mu \approx 10^{-3}$ Pa-s is the 
viscosity of water. If we take $v \sim 60$ kbp/s, the highest velocity indicated by the data, then 
$f_{\text{drag}} \sim 0.04$ pN may be regarded as the maximum frictional drag 
arising from the tail. This is much less than the $10-40$ pN driving 
force~\cite{grayson_real-time_2007} ejecting the DNA from the capsid and therefore cannot be the principal 
source of resistance.  Similarly, it is easy to show that viscous friction from the medium external to the capsid 
cannot be important either. The actual drag would depend on the shape of the ejected DNA coil, but, 
an estimate, that should be regarded as an upper bound, is obtained by regarding the entire $N_{max}=48.5$ kbp DNA to be a cylinder that 
translates in water parallel to its axis 
at a speed of $v \sim 60$ kbp/s. In this case, the drag force is~\cite{happel_brenner}
\begin{equation} 
f_{\text{drag}} = \frac{4 \pi \mu b N_{max}v}{\ln (bN_{max}/a) + \ln 2 - 1/2} \approx 0.4  \text{ pN},
\label{visc_drag_cyl1}
\end{equation} 
which is again negligible in comparison to the driving force.
However, the driving force does decay rapidly with $N$, 
and for $N < 5$ kbp the driving force is less than a pN. Thus, the hydrodynamic drag 
could balanced the driving force towards the very end of the translocation process, and 
 indeed, the data shows that $m(N)$  stops decreasing once $N$ becomes smaller than 
about 10 kbp (Fig.~\ref{fig:plot}).

Fig.~\ref{fig:plot} also shows that as long as $N > 10-20$ kbp, the experimental data can be fit very 
well by a function of the form 
\begin{equation} 
f_{\text{drag}}/v = m(N) = A \exp(k N).
\label{visc_drag_exp}
\end{equation} 
The fitted straight line corresponds to $A= 0.012$ pN-s/kbp and $k = 0.105$ kbp$^{-1}$. 
It is significant that $m(N)$ is hardly affected when the buffer contains 
the divalent Mg$^{2+}$ ion which reduces the electrostatic repulsion between the DNA strands thereby reducing the 
driving force by almost an order of magnitude. This is consistent with our expectations, since 
frictional resistance is not expected to be affected greatly by interstrand repulsion. 

The success of the empirical  fit, equation~(\ref{visc_drag_exp}), makes one wonder if there is any known 
physical mechanism that could lead to such an exponential friction law. We show here that if one assumes that the resistance 
to motion arises due to friction between the sliding DNA strand and its neighboring strands 
as well as with the capsid wall, then such an 
exponential law can be derived. For simplicity, we consider the shape of the viral capsid to be a cylinder,
though bacteriophage capsids are typically polyhedral in shape and are sometimes modeled as spheres.
We start from the equations of equilibrium of an elastic beam~\cite{landau_theory_1986}
\begin{eqnarray} 
\frac{d \mathbf{F}}{ds} &=& - \mathbf{L} \label{beam_force}\\
\frac{d \mathbf{M}}{ds} &=&  \mathbf{F} \times \mathbf{\hat{t}} \label{beam_moment}
\end{eqnarray}
where $s$ is the arc length along the beam, $\mathbf{F}$ is the elastic force across a cross-section,
$\mathbf{M}$ is the internal bending moment and $\mathbf{L}$ is the external force per unit length along the 
beam. Since the pitch of the helix is small~\cite{purohit_mechanics_2003}, we will neglect its torsion.
The external force is, by Amonton's law~\cite{bowden_friction_2001},  
$\mathbf{L} = - N  \mathbf{\hat{n}}  - f N \mathbf{\hat{t}}$,
where $\mathbf{\hat{t}}$ and $\mathbf{\hat{n}}$ are the unit tangent and normal to the curve representing the beam centerline, $\mathbf{\hat{t}}$ points in the direction of sliding 
and $f$ is a friction coefficient.
The force across the cross section may be resolved into a tensile ($T$) and shear ($S$) component:
$\mathbf{F} = T  \mathbf{\hat{t}} + S \mathbf{\hat{n}}$. The bending moment is related to the local curvature
($\kappa$) as $\mathbf{M} = EI \kappa \mathbf{\hat{b}}$, where $\mathbf{\hat{b}} = \mathbf{\hat{t}} \times \mathbf{\hat{n}}$,
$E$ is the Young's modulus and $I$ is the area moment of inertia of the cross-section.
Substitution in equations~(\ref{beam_force}) and (\ref{beam_moment}) and use of the 
Frenet-Serret formulas~\cite{struik_lectures_1961}
describing how the unit vectors $(\mathbf{\hat{t}}, \mathbf{\hat{n}}, \mathbf{\hat{b}})$
change along a curve in three dimensions results in 
\begin{eqnarray} 
\frac{dT}{ds} - \kappa S &=&  f N, \\
\frac{dS}{ds} + \kappa T &=&  N, \\
E I \frac{d \kappa}{ds} = -S.  
\end{eqnarray}
For a helix, $\kappa$ is a constant, thus $S=0$, and therefore,
\begin{equation} 
\frac{dT}{ds} = f \kappa T, 
\end{equation} 
which may be integrated to yield 
\begin{equation} 
T_{1} = T_{0} \exp ( f  \phi )
\label{capstan_equation}
\end{equation} 
where $T_{1}$ is the tension in the DNA at the capsid exit and $T_{0}$ is the tension at the terminal point.
By terminal point we mean the point on the DNA within the capsid marking the transition from a helix 
to the loosely coiled trailing end in the central core. If we express the viscous resistance on this trailing 
end as $v m_{0}$, then $T_{0}=vm_{0}$. Since the pitch of the helix is small, 
$\phi \approx 2 \pi b N /R$, so that 
\begin{equation} 
m(N) = m_{0} \exp (  2 \pi b f N / R),
\label{derivemN}
\end{equation} 
where $R \sim 25$ nm, is the capsid radius. It should be noted that in equation~(\ref{capstan_equation}),
$T_{1}/T_{0}$ is independent of the radius of the turns but depends only on the total turn angle, $\phi$. Thus, if there are multiple layers of coils, it may still be applied as long as 
the friction coefficient between all contacting surfaces are the same.
Equation~(\ref{derivemN}) is equivalent to equation~(\ref{visc_drag_exp});
comparing the coefficients we determine $f \approx 1.5$. 

It is interesting to note that equation~(\ref{capstan_equation}) is the capstan equation (also known as 
 Eytelwein's formula or the Euler-Eytelwein formula) familiar from mechanics~\cite{stuart_capstan_1961}.
 In the classical ``capstan problem'' a  flexible line is wound around a cylinder (for example, a bollard or capstan used in ship mooring)
to hold a load at one end of the line by the application of a holding force. At the point when the 
line is just about to slip, the relation between the load ($T_{1}$) and holding force ($T_{0}$) is described 
by equation~(\ref{capstan_equation}).
The exponential dependence of $T_{1}/T_{0}$ on  $\phi$ explains how a very large load can be supported 
by a relatively modest force if the rope is wound even a few turns around the capstan.
The name ``capstan friction model'' in the title of this paper 
is a reference to this correspondence. It should be noted, however, in our case 
there is no central cylinder. The frictional resistance arises from contact between neighboring 
DNA strands and possibly with the capsid wall as the tightly coiled DNA unwinds. In fact, the unfurling of 
a surveyor's steel measuring tape is perhaps a better analogy.

The frictional coefficient between a pair of surfaces depends on the nature of 
the contact between them~\cite{bowden_friction_2001,persson_sliding_2000}.
When a lubricating fluid film is present, the coefficient is small but increases linearly 
with sliding speed in a manner that may be well understood from the equations of fluid 
flow. In the presence of a large normal load (which in the present problem would 
arise from the strong bending rigidity of the DNA) the
fluid film thins and one enters the regime of boundary lubrication 
where direct molecular level  contact between surfaces could occur. This is characterized by a sharp rise 
in the frictional coefficient. The 
value $f \approx 1.5$ corresponds to this latter regime of boundary lubrication. 
For example, the frictional coefficient between dry highly polished metal surfaces at moderately 
large loads is close to this value~\cite{bowden_friction_2001}. The derivation 
of equation~(\ref{derivemN}) presumes Amonton's law 
of friction which is based on the idea that contact between a pair of surfaces 
arise from interlocking asperities and is thus independent of the apparent area of 
contact but increases in proportion to the normal load~\cite{bowden_friction_2001}. 
Such a picture is unlikely to hold for nanoscale phenomena such as the sliding of DNA strands. Nevertheless, 
Amonton's law is frequently used to describe frictional effects in nanoscale phenomena even though 
the underlying physical models are different from the rubbing of asperities envisaged 
to rationalize Amonton's law in the classical treatment of 
friction~\cite{bhushan_springer_2004,gao_frictional_2004,mo_friction_2009-1}. 

It may be shown by simple estimates that viscous friction with the fluid in the core is unimportant. 
Since any shear induced at the outer boundary of the fluid filled core will equilibrate on a 
time scale $R^{2}/\nu \sim 1$ ns ($\nu$ is the kinematic viscosity of water), we conclude, that,
 the fluid core must be in rigid rotation during most of the ejection process of duration $\tau \sim 1$ s.
The inertia of this fluid mass also plays no role. Indeed, the moment of inertia of the cylinder 
is $I \sim \rho R^{5}$ and the rate of change of angular speed is $\dot{\omega} \sim b N_{max} / R \tau^{2}$
where $N_{max}$ is the total length of DNA. 
Substituting typical values~\cite{grayson_real-time_2007} 
we find that $I \dot{\omega} \sim 10^{-32}$ N-m which is many orders of magnitude smaller
than the applied torque, $M \sim - R F^{\prime}(N)/b \sim 10^{-18}$ N-m. Thus, neither 
the viscosity of the water within the capsid nor  its inertia plays a significant role in determining the 
ejection velocity. 

In this paper we interpret  phenomena involving the translocation of single molecules 
of DNA across nanometer size pores from the principles of classical continuum mechanics. 
In the light of the small scale nature of the system, one might question the validity of such an approach. 
However, even though the relevant length scales are approaching the limits of applicability 
of the continuum description of matter, classical continuum mechanics has been applied successfully 
to other problems involving translocation of single molecules across 
nanopores~\cite{ghosal_electrophoresis_2006,ghosal_effect_2007,luan_electro-osmotic_2008,van_dorp_origin_2009}. 
Thus, its use in this context, for estimating forces on DNA, appears entirely reasonable.\\[2ex]

\noindent {\em Acknowledgement:} This
work was supported in part by the American Recovery and Reinvestment
Act (ARRA) funds through grant number R01HG004842 to Northwestern
University (USA) from the National Human Genome Research Institute,
National Institutes of Health. Support from the Leverhulme Trust (UK) in the form 
of a Visiting Professorship is gratefully acknowledged. The author thanks 
the Cavendish Laboratory, Cambridge University for its kind hospitality.
%\bibliographystyle{prsty}
%\bibliography{zotero_lib,microfluidics}

\end{document}